\documentclass[a4paper,scriptaddress,twocolumn, prd,showkeys]{revtex4-1}

\usepackage{amssymb} 
\usepackage{amsmath}
\usepackage{amsfonts}
\usepackage{mathrsfs} 
\usepackage{eulervm}

\makeindex

\newcommand{\bra}{\langle}
\newcommand{\ket}{\rangle}
\newcommand{\del}{\partial}
\newcommand{\rd}{\mathrm{d}}

\newcommand{\tr}{\mathrm{Tr}}

\begin{document}

\title{Identities for entropy change associated with time-evolution of an open system}
\author{Hiroki Majima}
\email{majima@salesio-sp.ac.jp}
\affiliation
{Department of General Education, Salesian Polytechnic \\
4-6-8 Oyamagaoka, Machida, Tokyo 194-0215, Japan}

\author{Akira Suzuki}
\affiliation
{Department of Physics, Faculty of Science, Tokyo University of Science \\
1-3 Kagurazaka, Shinjyuku-ku, Tokyo 162-8601, Japan}

\begin{abstract}
A general relation between entropy and an evolutionary superoperator is derived based on the theory of the real-time formulation. 
The formulation establishing the relation relies only on the framework of quantum statistical mechanics and the standard definition of the von Neumann entropy. 
Applying the theory of the imaginary- time formulation, a similar relation is obtained for the entropy change due to the change in reservoir temperatures. 
To show the usefulness of these formulas, we derived the expression for the entropy production induced by some dissipation in an open quantum system as the exemplary model system.
\end{abstract}


\maketitle



%
\section{Introduction}\label{sec1}
%
A remaining big challenge in modern physics is probably to understand universal aspects of nonequilibrium states in microscopic systems, and/or to develop nonequilibrium thermodynamics and statistical mechanics. 
Nonequilibrium phenomena are ubiquitous in nature. 
Yet, a general framework allowing their description far from equilibrium is lacking\,\cite{lebon}. 
A defining feature of out-of-equilibrium systems is that they dissipate energy in the form of heat, leading to an irreversible increase of their entropy. 
The nonequilibrium entropy production and its time derivative (the entropy production rate) are therefore two fundamental concepts of nonequilibrium thermodynamics\,\cite{lebon}. 
The subject of time-evolution of open systems with reference to the entropy production has been widely considered in the frame of quantum theory and quantum field theory, in particular, for example in the monumental work done by Umezawa and his school\,\cite{umezawa,takahashi,celeghini,blasone}.

Traditionally, the entropy production is used to evaluate the performance of thermodynamic devices: the maximal useful work, the availability (or energy), that can be extracted from a given system, is reduced by the presence of irreversibilities, such as friction or nonstatic transformations. 
This loss of availability is directly related to the mean entropy production\,\cite{cengel}.

In recent years significant insights into the nature of irreversibility and entropy production have emerged, partly due to the need for a framework to interpret thermodynamic processes on the nanoscale. 
In order to develop a theory for deterministic thermostatted systems, we take into account the dissipation function and fluctuation theorem\,\cite{evans1,evans2,evans3,carberry}.  
Some powerful formulas such as the Crooks and Jarzynski relations\,\cite{jarzynski,crooks1,crooks2} stand out along with a unifying framework for overdamped Langevin dynamics\,\cite{seifert} based on a stochastic description of the first law of thermodynamics commonly referred to as stochastic energetics\,\cite{sekimoto}.  
On the other hand, since the early ideas of Kirchoff, Rayleigh-Onsager and Prigogine\,\cite{kirchoff,onsager,prigogine}, it was hoped, thought, and believed that nonequilibrium regimes could be characterized by variational principles. 
Prigogine's minimum entropy production principle characterizes the nonequilibrium steady state as an extremum of the entropy production rate.
For typical example, the stationary state in heat conduction is characterized by a minimum of the entropy production, compatible with the imposed temperature distribution at the walls of the system\,\cite{deGroot}.
These works indicate importance of obtaining non-trivial relations in analyzing nonequilibrium systems.

Although various available viewpoints onto the irreversibility nature differ from each other, it is claimed that the central point of view is a mechanical, or possibly dynamical, quantity can be defined throughout behavior of the thermodynamic entropy. 
Indeed, the positive aspect of nonequilibrium entropy production is its experimental verification\,\cite{deffner,lucarini}.  
Therefore, it is very important to find out some non-trivial relations, especially the relations between entropy production and its associated evolutionary operator that describes the behavior of dissipative systems.

This paper is organize as follows.  
In Sec.\,\ref{sec2}, a general formula, which relates the entropy and an evolutionary superoperator for a dissipative system, is derived within the framework of statistical mechanics and the definition of von Neumann entropy in the real time formulation.  
In Sec.\,\ref{sec3} a general formula for the entropy change for deterministic thermostated systems is derived by introducing the general theory of an imaginary-time formulation.  In order to demonstrate a usefulness of the formulas obtained in this work, we consider a  model for a quantum dissipative system in Sec.\,\ref{sec4}.  
Concluding remarks are given in the final section.

%
\section{Real-time formulation}\label{sec2}
%
A general formula, which connects entropy with an evolutionary superoperator, is derived by using only an entropy operator and a causal evolution equation.  
In the microscopic picture, the state of a system is generally expressed by a density operator $\rho$. 
For every physical system there exists a superoperator ${\mathcal T}$ (the evolutionary superoperator), defined on the state space of the system, which determines the causal evolution of the state descriptor $\rho$ via the following law: 
\begin{align}
 \frac{\del\rho}{\del t}={\mathcal T}\rho\,,\label{eq1}
\end{align}
where the evolutionary superoperator ${\mathcal T}$ corresponds to the Liouvillian superoperator for an isolated system. 
The expression for the evolutionary superoperator ${\mathcal T}$ depends on the system considered.

Before introducing the explicit form of ${\mathcal T}$, let us first derive a general relation between entropy operator $\eta$ and the evolutionary superoperator ${\mathcal T}$.

The entropy operator is introduced as a dynamical variable in terms of the density operator $\rho$.  
The entropy operator $\eta$ is defined as~\cite{mori} 
\begin{align}
 \eta=-k\ln\rho\,, \label{eq2}
\end{align}
where the presence of $k$ (Boltzmann's constant) ensures the von Neumann entropy denoted by $S$. See below.

The expectation value of $\eta$ is then given by
\begin{align}
 \bra\eta\ket = \tr\,\rho\,\eta = -k\,\tr\,\rho\ln\rho\equiv S\,, \label{eq3}
\end{align}
where the statistical average of $\eta$ denoted by a symbol $\bra\eta\ket$ is the well-known ``von Neumann entropy" if $k$ is taken as the Boltzmann constant.  
We call the $S$ defined in Eq.\,(\ref{eq3}) ``entropy" hereafter.  
From Eq.\,(\ref{eq2}), the density operator $\rho$ for the system  may be expressed in terms of the entropy operator $\eta$:
\begin{align}
 \rho=e^{-\eta/k}\,, \label{eq4}
\end{align}
where the time dependency of the density operator $\rho$ comes from the entropy operator $\eta$.   
It should be noted that $\eta$ is indeed a function of $\rho=\rho(t)$, {\it i.e.}, $\eta(t)=\eta[\rho(t)]$\,\cite{callen}.

Taking the time derivative of Eq.\,(\ref{eq4}), we obtain
\begin{align}
 \frac{\del\rho}{\del t} 
 =-\frac{\rho}{k} \int_0^1K(\lambda)\rd\lambda\,,  \label{eq5}
\end{align}
where $K(\lambda)$ is given by 
\begin{align}
 K(\lambda)
 &\equiv 
 e^{\lambda\eta/k}\,\frac{\del\eta}{\del t}\,e^{-\lambda\eta/k}\nonumber\\
 &=\frac{\del\eta}{\del t}
   +\lambda\Bigl[\frac{\del\eta}{\del t},\frac{\eta}{k}\Bigr]
   +\frac{\lambda^2}{2!}
     \biggl[\Bigl[\frac{\del\eta}{\del t},\frac{\eta}{k}\Bigr],\frac{\eta}{k}\biggr]
   +O(\lambda^3)\,. \label{eq6}
\end{align}
Comparing Eq.\,(\ref{eq1}) with Eq.\,(\ref{eq5}), we obtain 
\begin{align}
\rho \int_0^1K(\lambda)\,\rd\lambda 
 =-k\,{\mathcal T}\rho\,. \label{eq7}
\end{align}
This is an {\it exact} operator identity.  
Taking a trace of Eq.\,(\ref{eq7}), we can write 
\begin{align}
\Big\bra\int_0^1K(\lambda)\rd\lambda\Big\ket
=-k\,\tr\,{\mathcal T}\rho\,.\label{eq8}
\end{align}
This formula connects the entropy change rate $\rd S/\rd t$ to the evolutionary superoperator ${\mathcal T}$ of the system under study.

The equations (7) and (8) were strictly obtained from the causal evolution of $\rho$ and the definition of the entropy operator $\eta$ only. 
Noting that
\begin{align}
\frac{\rd S}{\rd t}=\frac{\rd\bra\eta\ket}{\rd t}
={\Bigl\bra}\frac{\del\eta}{\del t}{\Bigr\ket}\,, \label{eq9}
\end{align}
we can obtain entropy $S$ from Eq.\,(\ref{eq8}). 
The last equality in Eq.\,(\ref{eq9}) holds since the trace operation is time independent, i.e.,    
$\bra\del\eta/\del t\ket=\rd\bra\eta\ket/\rd t$. 
The similar relation was originally derived in the framework of 
thermofield dynamics by the present authors~\cite{majima1,majima2,majima3}.

Using Eqs.\,(\ref{eq8}) and (\ref{eq9}), the entropy change $\Delta S$ can be obtained in terms of the evolutionary superoperator ${\mathcal T}$ in the following form:
\begin{align}
 \Delta S =-k\int\rd t\,\tr\,{\mathcal T}\rho\,. \label{eq10}
\end{align}
The relations (\ref{eq8}) and  (\ref{eq9}) were derived by using the causal evolution equation (\ref{eq1}) and the defining equation (\ref{eq2}) for the entropy operator $\eta$.

The entropy change $\Delta S$ induced by, e.g., the spontaneous or irreversible processes, can be evaluated by using the formula (\ref{eq10}), which relates the entropy $S$ to an evolutionary superoperator ${\mathcal T}$ associated to the system.  
Thus by using the obtained formula, the effect of time-evolution of the system due to, e.g., the spontaneous or irreversible process of the system can be theoretically investigated by giving the explicit form of the evolutionary superoperator if the density operator $\rho$ is given.

Implementing Eq.\,(\ref{eq9}) into Eq.\,(\ref{eq4}), we receive a nonequilibrium distribution function at a time $t$ in terms of the entropy $S$ of the system: 
\begin{eqnarray}
f(t)&=&\exp({-S/k})
=C\exp\Bigl(
  -\frac{1}{k}\int\rd t\,\Bigl\bra\frac{\del\eta}{\del t}\Bigr\ket\Bigr) \nonumber\\ 
  &=& C\exp\int\rd t \,\tr\,{\mathcal T}\rho ,  \label{eq11}
\end{eqnarray}
where $C$ is a normalization factor. 
If we adopt the initial condition $f(t_0)=Z^{-1}e^{-\beta E}$ at $t=t_0$, Eq.\,(\ref{eq11}) can be expressed by 
\begin{align}
 f(t)=Z^{-1}e^{-\beta E}\exp\int_{t_0}^{t}\rd t\,\tr\,{\mathcal T}\rho\,,\label{eq12}
\end{align}
where $Z$ is the canonical partition function and $E$ is the energy of the system at $t=t_0$.

%
\section{Imaginary-time formulation}\label{sec3}
%
In the imaginary time formulation (ITF), the evaluation of thermal averages is performed by replacing the time t by the imaginary time $i\beta$, according to the formal analogy between the imaginary time and the inverse temperature first noticed by F. Bloch \cite{bloch}.
Time integration is restricted to a finite domain along the imaginary axis, from $0$ to $i\beta$. 
Thus, the time evolution in the real time formulation corresponds to the temperature evolution in the ITF.

Let us first derive a general formula for the entropy production induced by the change in reservoir temperatures. 
In order to treat such a system, it is convenient to rewrite the formulas obtained in Sec.\,\ref{sec2} by replacing the imaginary time $it$ by temperature $\beta$ ({\it i.e.}, $it\to\beta$).  
Here we regard the imaginary time $it$ as temperature $\beta\,\, (\equiv 1/kT)$, and carry out  the Wick rotation of the imaginary time variable.  
The causal evolution equation (\ref{eq1}) can then be expressed in terms of $\beta$:
\begin{align}
 \frac{\del\rho}{\del\beta}=-i\,{\mathcal T}\rho\,, \label{eq13}
\end{align}
and Eq.\,(\ref{eq5}) in the Wick rotated frame is given by
\begin{align}
 \frac{\del\rho}{\del\beta}
 =\frac{i\rho}{k}\,\int_0^1K_{\beta}(\lambda)\,\rd\lambda\,, \label{eq14}
\end{align}
where $K_{\beta}(\lambda)$ is given by
\begin{align}
 K_{\beta}(\lambda)
 \equiv 
 ie^{\lambda\eta/k}\,\frac{\del\eta}{\del\beta}\,e^{-\lambda\eta/k}\,. \label{eq15}
\end{align}
Equating Eq.\,(\ref{eq13}) to Eq.\,(\ref{eq14}), we obtain the exact expression:
\begin{align}
\rho\int_0^1K_\beta(\lambda)\,\rd\lambda
=-k\,{\mathcal T}\rho\,. \label{eq16}
\end{align}

In the same way to derive the formula (\ref{eq9}), we formally expand $K_\beta(\lambda)$ with respect to $\lambda$ and then take its trace, and we can obtain the general formula:
\begin{align}
 \Bigl\bra\frac{\del\eta}{\del\beta}\Bigr\ket
 =ik\,\tr\,({\mathcal T}\rho)\,.\label{eq17}
\end{align}

This equation relates the evolutionary superoperator ${\mathcal T}$ to the entropy change  induced by the change of temperature $T$ if $\beta=(kT)^{-1}$, where $k$ is the Boltzmann constant. 
We can obtain the entropy change $\Delta S$ induced by the the temperature change in the following form:
\begin{align}
\Delta S= ik\int\rd\beta{\mathrm{Tr}}\,{\mathcal T}\rho\,.\label{eq17a}
\end{align}

Now we consider a canonical ensemble.  
In thermal equilibrium, the density operator is given by $\rho\sim e^{-\beta H}$.  
This unnormalized density operator satisfies the equation
\begin{align}
 \frac{\del\rho}{\del\beta}=-H\rho\,, \label{eq18}
\end{align}
where $H$ denotes the Hamiltonian of the system at the temperature $T\equiv 1/(k\beta)$. 
Equation (\ref{eq18}) is referred to as the Bloch equation for the density operator of a canonical ensemble.

Applying Eq.\,(\ref{eq18}) instead of  Eq.\,(\ref{eq13}), Eq.\,(\ref{eq17}) can be expressed and related to the entropy $S$ by
\begin{align}
\frac{\rd S}{\rd\beta}= \Bigl\bra\frac{\del\eta}{\del\beta}\Bigr\ket
 =k\bra H\ket\,, \label{eq19}
\end{align}
where $H$ denotes the Hamiltonian.

Hence, the generalized thermal distribution function $f(\beta)$ is thus obtained in a similar manner as in the real-time formulation by making use of Eq.\.(\ref{eq18}):
\begin{align}
 f(\beta)
 &=\,\exp(-S/k)
 =\,C\exp\Bigl(
    {-\frac{1}{k}\int\rd\beta\Bigl\bra
      \frac{\del\eta}{\del\beta}
      \Bigr\ket_{\hspace{-1mm}}}
   ~\Bigr)
  \nonumber\\
 &=\,C\exp
   \Bigl[-\int\rd\beta
       \bra H\ket
   \Bigr]\,,\label{eq20}
\end{align}
where $C$ is a normalization factor.  
If we assume the initial condition $f(\beta_0)=Z^{-1}e^{-\beta_0E}$ at temperature $T_0=(k\beta_0)^{-1}$, Eq. (\ref{eq20}) can be expressed by 
\begin{align}
 f(\beta) 
 =Z^{-1}e^{-\beta_0E}
   \exp
   \Bigl[-\int_{\beta_0}^{\beta}\rd\beta
       \bra H\ket
   \Bigr]\,,\label{eq21}
\end{align}
where $Z$ is the canonical partition function and $E$ is the energy of the system at the initial reservoir temperature $T_0$.

%
\section{Quantum dissipative system}\label{sec4}
%
Now, we apply the formula (\ref{eq9}) obtained in Section\,\ref{sec2} to an open quantum system with dissipation and consider how entropy is affected by the evolutionary superoperator associated to the open quantum system.

A quantum mechanical transition between the states of the system will be considered. 
In the case where final states are continuous, we can use Weisskopf-Wigner approximation\,\cite{Wigner}. 
The Hamiltonian of such systems can be generally expressed by 
\begin{align}
 H=H_0+i\varGamma, \label{eq22}
\end{align}
where $H_0$ and $\varGamma$ are Hermitian operators, denoting the Hamiltonian of an open system and the system-reservoir interaction, respectively.

In this model, the time-evolution of the density operator follows the Liouville-von Neumann equation: 
\begin{eqnarray}
 \frac{\del\rho}{\del t}
 &=&-\frac{i}{\hbar}(H\rho-\rho H^{\dag})\nonumber\\ 
 &=&-\frac{i}{\hbar}[H_0,\rho]
    +\frac{1}{\hbar}\{\varGamma,\rho\}
 \equiv {\mathcal T}\rho\,, \label{eq23}
\end{eqnarray}
where the curly brackets indicate the anticommutator defined for any operators $A$ and $B$: $\{A,B\}\equiv AB+BA$. 
If $\rho$ commutes with $H_0$ (i.e., $[\rho, H_0]=0$), the entropy production rate can then be obtained from the formula (\ref{eq9}) as in the form:
\begin{align}
\frac{\rd S}{\rd t}= \Bigl\bra\frac{\del\eta}{\del t}\Bigr\ket
 =-\frac{2k}{\hbar}\tr\,\varGamma\rho\,, \label{eq24}
\end{align}
where ${\varGamma}$ is an operator representing dissipation (damping).  
We can evaluate the entropy production rate by giving a specific form (microscopic description) of ${\varGamma}$ and thus obtain the ${\varGamma}$-related information  theoretically.  
The obtained formula enables us to obtain some information about the entropy directly from the evolutionary superoperator ${\mathcal T}$ in the equation of motion for the density operator $\rho$.

%
\section{Concluding remarks} \label{sec5}
%
A microscopic theory of entropy was developed and a general formula connecting entropy to the evolutionary superoperator associated with a system was derived.  
We also derived some general relations by making use of {\it only} the causal evolution of the density operator {\it and} the definition of the von Neumann entropy in the framework of quantum statistical mechanics.  
Making use of the relations, it is easy to obtain the entropy change rate {\it directly} from the evolutionary superoperator ${\mathcal T}$. 
Also we showed that by making use of these relations, the nonequilibrium distribution function for the system can be derived by taking the expectation value of the entropy production, see Eq.\,(\ref{eq11}) in the real time formulation.
It is worth stressing that that the distribution function has the form which is not an expectation value of exponential function but an exponential function of the expectation value.
Consequently, we have obtained a specific form of the distribution function of the concerned system.

In the imaginary-time formulation we obtained the relation (20) from the Bloch equation and the statistical entropy in a similar manner to the real-time formulation discussed in Sec. 2.
We also derived the generalized thermal equilibrium distribution function from Eq.\,(\ref{eq20}) in the imaginary time formulation. 
The distribution function in imaginary-time formulation recover the conventional thermal equilibrium distribution function by taking the lowest-order approximation for the entropy production.

The relations obtained in Sec. 2 can be applied to an open quantum system. 
In order to show the applicability of obtained formulas, we considered the case where the interaction causes dissipation. 
We obtained the formula for the entropy production due to the dissipation arising from a specific interaction by giving the specific interaction Hamiltonian. 
In other words, if we find the microscopic description of the interaction Hamiltonian ${\varGamma}$, all we need to do is to utilize the relations (\ref{eq8}) and (\ref{eq10}), and we will have the entropy production from which we can obtain some information about the interaction through the entropy of the system for the open quantum system.  
In the standard approach, entropy production is achieved through nonequilibrium thermodynamics.  
In our theory, however, we can {\it directly} evaluate entropy production by employing the formulas given in this paper.

Finally it is worth mentioning that a possible extension of our work would be a construction of the microscopic foundation of nonequilibrium thermodynamics.   
In particular our results would have the possibility to obtain microscopic representations of the system-reservoir interactions {\it i.e.} heat. 
Our new formulation presented in this article enables us to calculate the entropy production directly from system-reservoir (environment) interactions. 
This means that it allows us to conjecture what type of interactions play an essential role for frictional effects on the system through application of a conjectured form (microscopic description) of the interaction Hamiltonian ${\varGamma}$.
It goes beyond the standard derivation of the second law of thermodynamics.

%
%


\end{document}